# Analysis of Empirical Mode Decomposition-based Load and Renewable Time Series Forecasting

N. Safari, *Member, IEEE*, G.C.D. Price, and C.Y. Chung, *Fellow, IEEE*

*Abstract*— The empirical mode decomposition (EMD) method and its variants have been extensively employed in the load and renewable forecasting literature. Using this multiresolution decomposition, time series (TS) related to the historical load and renewable generation are decomposed into several intrinsic mode functions (IMFs), which are less non-stationary and non-linear. As such, the prediction of the components can theoretically be carried out with notably higher precision. The EMD method is prone to several issues, including modal aliasing and boundary effect problems, but the TS decomposition-based load and renewable generation forecasting literature primarily focuses on comparing the performance of different decomposition approaches from the forecast accuracy standpoint; as a result, these problems have rarely been scrutinized. Underestimating these issues can lead to poor performance of the forecast model in real-time applications. This paper examines these issues and their importance in the model development stage. Using real-world data, EMD-based models are presented, and the impact of the boundary effect is illustrated.

## I. Introduction

WITH increased demand response and renewable energy penetration, accurate load and renewable generation forecasting is of increased importance for optimal power systems operation. In this regard and with the aim of achieving high accuracy forecasts, tremendous efforts have been devoted to the development of both load and renewable forecast models utilizing state-of-the-art signal processing and machine learning techniques (e.g., [1,2]).

Due to the non-linear and non-stationary nature of load and renewable generation time series (TS), signal processing approaches have been employed to decompose the original TS into more predictable components [3]. Amongst decomposition approaches, empirical mode decomposition (EMD) and its variants (e.g., [4-9]) have attracted special attention with respect to load and renewable generation forecasting applications [10,11]. Compared to conventional decomposition approaches (i.e., Fourier analysis, wavelet decomposition, etc.), EMD is capable of intuitively decomposing a non-stationary and non-linear TS without *a priori* knowledge of the TS [4].

EMD is an adaptive time-frequency data analysis tool used for a wide range of applications. The prevalent EMD is subjected to modal aliasing in the presence of intermittencies and abnormalities in the TS [5]. As a result, applying EMD to the TS may lead to intrinsic mode functions (IMFs) without physical meaning. To alleviate this downside, variants of EMD have been put forward. In [5], a noise-assisted data analysis method known as ensemble EMD (EEMD) is proposed. EEMD has created new challenges, including contamination of the IMFs with noise, which are addressed in the more recent EEMD variants such as complementary EEMD (CEEMD) [7], complete ensemble empirical mode decomposition with adaptive noise (CEEMDAN) [8], improved complete ensemble empirical mode decomposition with adaptive noise (ICEEMDAN) [9], etc.

Due to these advancements in EMD techniques, many researchers have developed EMD-based forecasting models for load and renewable generation. For instance, Ref. [10] employs EMD to decompose the load TS into several IMFs, which are categorized into low- and high-frequency components. The aggregation of low-frequency components, representing the cyclic trend, is forecasted using a recursive autoregressive support vector regression (SVR), while the remaining components are approximated using the periodicity information and a non-recursive SVR. Authors in [12,13] use EMD variants in load forecasting models in which every IMF is forecasted independently. In [3,14], the TS associated with the IMFs and their lagged forms are utilized to construct a feature matrix that is subsequently used as the input of a single SVR for wind speed forecasting. Ref. [3] compares the performance of various EMD variants in wind speed forecasting. In [15], two wind speed TS forecasting models are presented in which the TS is decomposed using EMD and EEMD, and then extreme learning machine (ELM) is trained to forecast every IMF separately. The application of ICEEMDAN for wind power forecasting is demonstrated in [2].

Despite the promising forecasting accuracy reported in the literature, EMD-based forecasting tools are prone to various issues in the real-time application, so these must be considered in the model development stage. The boundary effect issue [16-

This work was supported in part by the Natural Sciences and Engineering Research Council of Canada and in part by SPC.

N. Safari, G.C.D. Price are with the Grid Operations Support, SaskPower, Regina, SK S4P 0S1, Canada (e-mail: nsafari@saskpower.com).

C.Y. Chung is with the Department of Electrical and Computer Engineering, University of Saskatchewan, Saskatoon, SK S7N 5A9, Canada.

20]—also known as the edge effect, endpoint effect, or end effect issue—is one of the most critical drawbacks of EMD-based models and has not been sufficiently considered in the development of EMD-based forecasting models. The authors of [21] present two EMD variants to address the boundary effect issue; however, as stated in the article, the proposed approach was unsuccessful.

To fill the gap in the literature, the impact of various EMD techniques on the TS analysis of load and renewable generation will be assessed herein for the first time. Moreover, the boundary effect problem in load and renewable generation TS forecasting is considered using real-world data and simulating real-time situations.

The remainder of this paper is organized as follows. Section II describes the data that are used throughout this paper. In Section III, the variants of EMD methods in load and renewable generation TS analysis are described, and the boundary effect problem is highlighted. Section IV summarizes the case studies and presents comparisons, and Section V concludes the paper.

## II. DATA DESCRIPTION

SaskPower is the power utility for the Province of Saskatchewan, Canada. Throughout this study, SaskPower hourly averaged net system load for January 2015-January 2016 has been used as the historical load data. Due to the non-disclosure agreement, the load data is shown on a normalized scale between -1 and 1. However, in the calculations of the forecasting accuracy, the data in the actual scales are used. Also, as a representation of the renewable generation dataset, the historical hourly averaged wind power generation data related to the Summerberry wind farm for February 2019-Februaty 2020 is used. This wind farm has a nameplate capacity of 20 MW, which is in the southeast of the province.

## III. DESCRIPTIONS OF EMD VARIANTS AND THEIR CHALLENGES

To better understand the potential issues of EMD-based forecasting models, this section briefly summarizes the prevalent EMD variants. A detailed description of the EMD method and its modified varieties can be found in [4-9].

### A. EMD Variants

Assume $X = \{X_1, ..., X_N\}$ is the load or renewable generation TS of interest, where $N$ is the total number of available historical sample points. The EMD steps are as follows:

**Step 1:** Find all local minimum and maximum points.
**Step 2:** Perform cubic spline interpolation using the minimum and maximum points to attain a lower envelope ($LE = \{LE_1, ..., LE_N\}$) and upper envelope ($UE = \{UE_1, ..., UE_N\}$), respectively.
**Step 3:** Find the difference between $X$ and the local average of $LE$ and $UE$ as follows:
$$h_n = X_n - \left(\frac{LE_n + UE_n}{2}\right) \quad (1)$$
**Step 4:** Repeat Steps 1-3 until $\left(\frac{LE_n + UE_n}{2}\right) \leq \epsilon$ for the final $h = \{h_1, ..., h_N\}$, where $\epsilon$ should be set very close to zero. The iterative procedure of Steps 2-3 is also known as the sifting process. As an alternative to the above-mentioned stopping criterion, the maximum number of repetitions in the sifting process may also be used.

**Step 5:** Consider the last $h$ as the IMF, $imf_i$, where $i$ is the IMF number, and $i = 1$ for the first IMF.
**Step 6:** Replace $X$ with the residual $X_n^{new}$ as follows:
$$X_n^{new} = X_n - imf_{i,n} \quad (2)$$
where $imf_{i,n}$ is the $n$th sample point of the $i$th IMF (i.e., $imf_i$).
**Step 7:** Repeat Steps 1-6 until all of the predefined number of IMFs, denoted by $IMF$, have been derived or the residual $X_n^{new}$ is monotonic. The maximum number of IMFs is limited to $\log_2 N$. The last $X_n^{new}$ is considered as the residual component, $r$.

The original TS can be reconstructed as follows:
$$\hat{X} = \sum_{i=1}^{IMF} imf_i + r \quad (3)$$
where $\hat{X}$ is the reconstructed TS, which theoretically should be the same as $X$. Please note executing Steps 1-7 might result in some of the IMFs being nonorthogonal with respect to each other, and amalgamation to construct a single component needs to be considered [5].

The described steps result in EMD being very sensitive to local extremum points, which can be affected if the TS is contaminated by noise-like components. Instances may occur when a specific IMF has sub-components with instantaneous frequencies that belong to other IMFs; alternatively, a specific IMF might not have sub-components with instantaneous frequencies, but the existence of these components is manifest in most of the IMFs. Load and renewable generation TS have noticeable amounts of high-frequency noise-like components as a result of the intermittent nature of the non-conforming loads and weather-related factors. To this end, EEMD-based methods [5], in which this problem has been addressed to a great extent, have been employed for load and renewable generation TS forecasting. The EEMD approach can be summarized in the following steps.

**Step 1:** Generate a white noise TS and add it to the original TS.
**Step 2:** Apply EMD to the obtained TS.
**Step 3:** Repeat Steps 1-2, $NE$ times, where $NE$ is the number of ensembles.
**Step 4:** Extract the final IMFs by averaging the IMFs of $NE$ ensembles.

As an example, Figs. 1 and 2 presents the discrete-time short-term Fourier transform (STFT) [22] for the fourth IMF of the load TS extracted by applying EMD and EEMD, respectively. A comparison of these figures shows that EMD results in an IMF with more local distortion. Few time instances have high-frequency components (i.e., approximate-ly greater than 0.15 cycles/hour) in the EMD-based IMF (Fig. 1). Between July and October, the frequency contents of the EMD-based IMF are distorted; however, the EEMD-based IMF shows a generally uniform frequency content throughout the study period. Therefore, EEMD is expected to provide more useful information about the trend and cyclic behavior of the TS. In other words, the lags of the IMF TS are expected to be more correlated with each other compared to the EMD-based IMF; this can be observed from the sample partial autocorrelation function (PCAF) shown in Fig. 3.

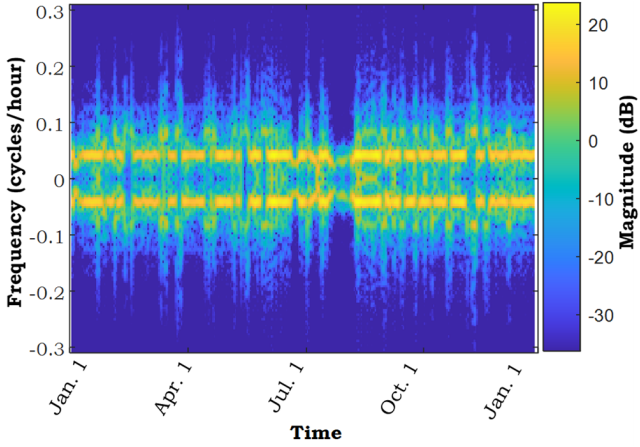

Fig. 1. STFT of the fourth IMF of the normalized load TS using EMD.

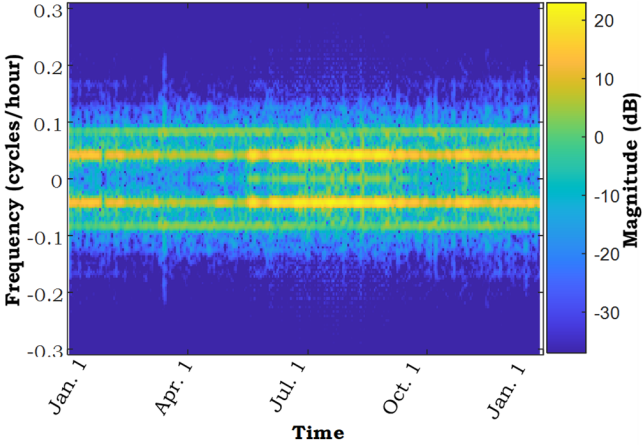

Fig. 2. STFT of the fourth IMF of the normalized load TS using EEMD.

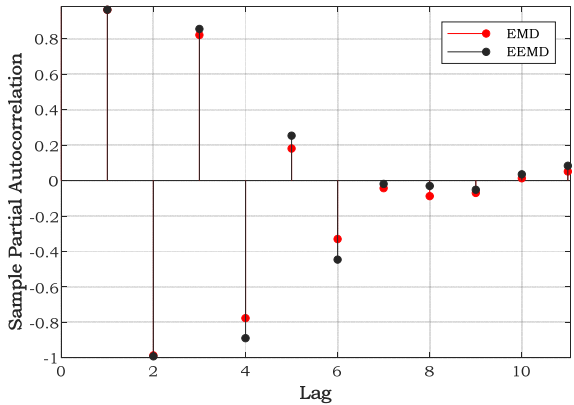

Fig. 3. Sample PCAF of the fourth IMF of EMD and EEMD, related to Figs. 1-2.

Combining the TS with white noise may result in residual noise if the number of ensembles is not adequately high. However, a large number of ensembles may be computationally intractable. To alleviate this issue, another variant of EEMD, known as CEEMD was proposed in [7]. The main difference between EEMD and CEEMD is the way that white noise is added to the original TS; in CEEMD, the TS is contaminated by white noise in pairs.

In other words, if in one ensemble white noise $\epsilon_n$ is added to the TS then in the subsequent ensemble the white noise $-\epsilon_n$ is added to the TS. To see the impact of CEEMD, the wind power TS described in Section II for the period of 2019-01-01 00:00 to 2019-02-01 00:00 is decomposed by both EEMD and CEEMD with 200 ensembles. The improvement of TS decomposition using CEEMD is assessed by measuring the signal-to-noise ratio ($SNR$), which is calculated as follows:

$$SNR = 10 \log_{10} \left( \frac{\sum_{n=1}^{N}(X_n)^2}{\sum_{n=1}^{N}(X_n - \hat{X}_n)^2} \right) \quad (4)$$

The $SNR$ values for CEEMD and EEMD are about 305 and 53, respectively. This example shows the noteworthy superiority of CEEMD compared to EEMD in terms of reconstructing the TS from the EEMD components.

Other variants of CEEMD, including CEEMDAN [8] and ICEEMDAN [9], have also been proposed to address the EEMD issues that result from the added noise.

### B. Boundary Effect

The boundary effect is a crucial drawback of the EMD-based decomposition. As explained in Section III-A, finding the local extremum points is the basis of the EMD method and its variants. At the boundary of the TS (i.e., in the vicinity of the last sample point), the determination of local extremum points requires assumptions as the next sample points are unknown. Many approaches have been proposed to circumvent this issue in various fields [17-20]; however, the importance of this issue in the load and renewable generation forecasting literature has been underestimated to a great extent.

To further clarify this issue in load and renewable generation, consider Fig. 4, which shows the IMFs estimations of the two days wind power generation TS from 2019-03-30 00:00 to 2019-04-01 00:00. As one can see, there are two estimations for each of the IMFs. In the first estimation, which is represented by a solid black line, no prior knowledge about the wind patter after 2019-04-01 00:00 was available. However, in the second estimation, shown by the dashed red line, the sample of the wind power TS after 2019-04-01 00:00 was known during the decomposition process. Therefore, the local minimum and maximum points at the boundary of the time window (i.e., 2019-04-01 00:00) is correctly identified. With no knowledge about the wind pattern at the boundary of the TS, linear extrapolation is used. The local extremum points in the vicinity of the endpoints are then identified. Fig. 4 vividly shows the imprecise extrapolation leads to erroneous detection of local extremum points; therefore, the pattern of IMFs at the boundary of the TS (i.e., sample point 48) diverges from the actual pattern. The boundary effect problem is exacerbated as the number of IMFs increases. By increasing the number of IMFs, the significant deviation of IMFs obtained from EMD without any prior knowledge of the TS values after 2019-04-01 00:00 and from EMD with prior knowledge of the TS values propagates to sample points far from the boundary of the TS.

Notably, the boundary effect problem exists at both ends of the TS. However, forecasting applications usually have sufficient historical data points such that the boundary effect at the beginning of the TS can be addressed by providing *a priori* knowledge about the data points collected before the start time of the decomposition. Hence, the boundary effect at the boundary corresponding to the end time is the most critical issue in EMD-based forecasting problems. In most of the

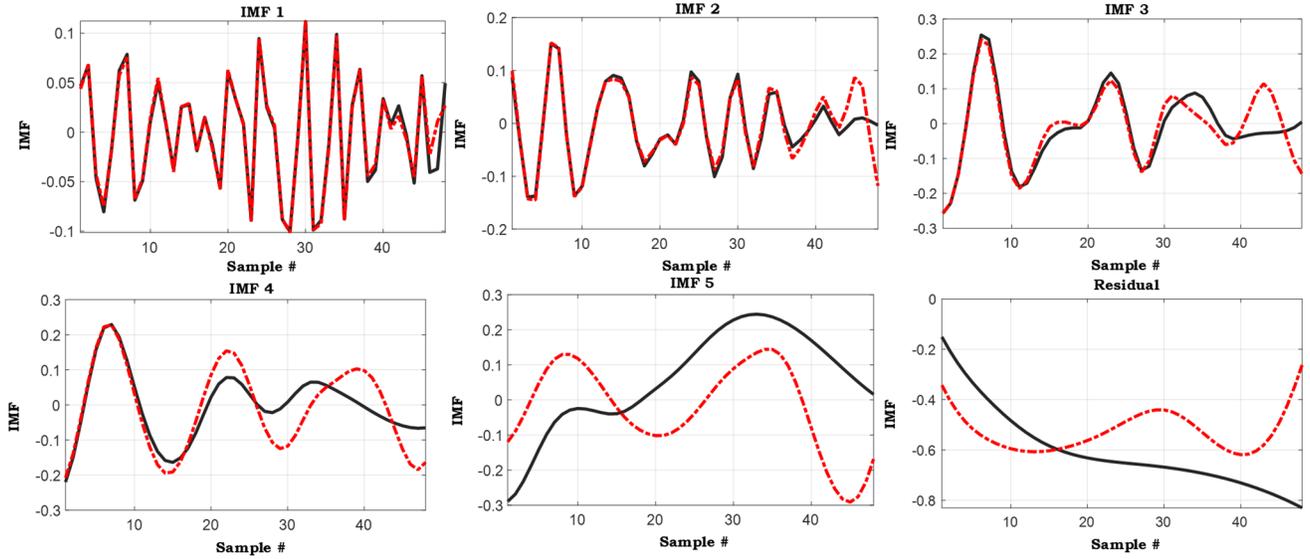
Fig. 4. Demonstration of boundary effect problem by comparing EMD with and without prior knowledge about the values of the TS beyond its boundaries.

EMD-based load and renewable generation forecast literature, the boundary effect is not taken into consideration either during model development or testing [21].

The importance of considering the boundary effect in load and renewable energy generation forecasting accuracy is empirically analyzed in Section IV.

## IV. CASE STUDIES

### A. Model and Scenario Description

Extreme learning machine (ELM) [15,16] and support vector regression (SVR) [13,14] are two of the most popular prediction engines, used in the EMD-based TS forecasting models. In the ELM-based load and renewable generation forecasting, a sigmoidal activation function has been used (e.g., [23]). While the radial basis function (RBF) is recommended as the kernel function for load and renewable generation forecasting (e.g., [13]). Gird search with five-fold cross-validation is utilized for choosing the suitable number of hidden neurons in ELM and kernel function parameter as well as the penalty factor in the SVM.

The flowchart of the 1-hour ahead prediction model, used in this study, is depicted in Fig. 5, which shows the TS is decomposed into $IMF$ components, and a prediction engine (i.e., ELM or SVR) is trained for forecasting each of the components. Finally, using (3) the TS is reconstructed; thus, the TS value for the next sample time is forecasted. Minimum redundancy and maximum relevance (mRMR) [24]—a widely used feature selection—is adapted for the selection of the informative lags as the predictors. The test model was designed to assess the performance of EMD-based forecasting models, rather than to develop a comprehensive model in which several informative external predictors are employed. Therefore, other predictors may also be considered to enhance the accuracy of the forecast, but this is outside the scope of this paper. Note that 80% of the data are used for training and validation and the remaining 20% for testing.

Scenarios I-III are considered to evaluate the impact of

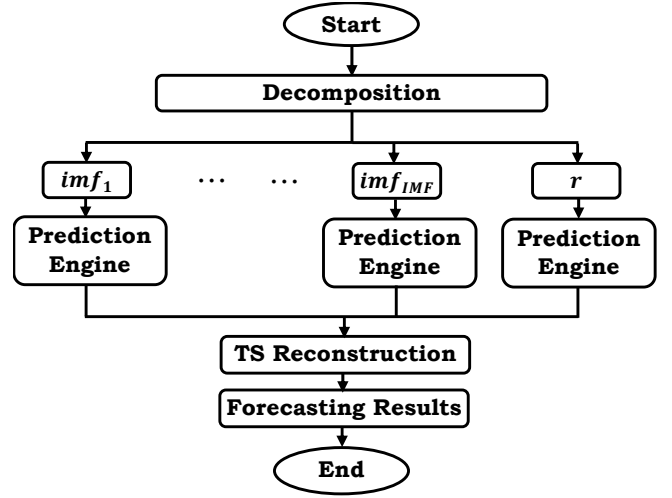
Fig. 5. Flowchart of the EMD-based forecasting model, used in the case studies.

boundary effects with respect to deterioration of forecast accuracy. In Scenario I, the original TS is decomposed using CEEMD, and the IMFs are then partitioned into training and testing. In Scenario II, only the training portion of the original TS is decomposed in the first step. After the model is trained, the real-time situation is simulated by adding the testing points one at a time, decomposing the TS, and feeding the testing input vector to the trained model. The model is updated as each new historical data point is added. Therefore, Scenario II represents the real-time application of EMD-based forecasting. In Scenario III, no decomposition is done.

### B. Evaluation Metrics

The widely used evaluation metrics of mean absolute percentage error ($MAPE$) [10] and root-mean-square error ($RMSE$) [3] are employed to assess the performance of the prediction models. $MAPE$ and $RMSE$ can be calculated as follows:

$$MAPE = \frac{1}{L}\sum_{l=1}^{L}\frac{|\hat{y}_l - y_l|}{y_l} \times 100 \quad (5)$$

$$RMSE = \sqrt{\frac{\sum_{l=1}^{L}(\hat{y}_l - y_l)^2}{L}} \quad (6)$$

TABLE I
EVALUATION RESULTS OF LOAD FORECASTING FOR DIFFERENT SCENARIOS

| Prediction Engine | Scenario | Testing | |
|---|---|---|---|
| | | MAPE | RMSE |
| ELM | I | 0.5800 | 21.6311 |
| | II | 3.2072 | 120.4768 |
| | III | 1.1841 | 43.9851 |
| SVM | I | 0.6754 | 24.7423 |
| | II | 2.7671 | 105.7308 |
| | III | 1.1190 | 43.8142 |

TABLE II
EVALUATION RESULTS OF WIND FORECASTING FOR DIFFERENT SCENARIOS

| Prediction Engine | Scenario | RMSE |
|---|---|---|
| ELM | I | 0.8757 |
| | II | 3.1836 |
| | III | 2.2541 |
| SVM | I | 0.9293 |
| | II | 3.2421 |
| | III | 2.1796 |

where $L$ is the number of testing points. In (5)-(6), $y_l$ and $\hat{y}_l$ are the actual and forecasted values of the $l$ testing point, respectively. As in the wind forecasting case study $y_l$ can be zero and consequently the denominator in the Eq. (5) becomes zero, MAPE is only used for load forecasting case study.

### C. Load Forecasting

The forecast error (i.e., MAPE and RMSE) for the 1-hour ahead prediction of load TS is summarized in Table I. Scenario I demonstrates the best performance in both ELM- and SVR-based models. This table also shows Scenario II, in which the decomposition has been performed on a real-time basis, results in poor forecast accuracy.

This case study reveals prediction engine type may slightly impact the degree of forecast accuracy. However, regardless of prediction engine type, Scenario II has inferior performance compared to Scenario III, in which no decomposition approach is employed. Therefore, in the case of using EMD variants in the load forecasting applications, tailored boundary reduction approaches are required.

### D. Wind Power Forecasting

Table II presents the RMSE related to wind power prediction in the various scenarios and using different prediction engines. Similar to the load forecasting case study, Scenario I outperforms the other scenarios irrespective of prediction engine, while Scenario II has the worst performance. In this regard, the results indicate the boundary effort is a critical issue in EMD-based wind power forecasting and should be considered in the model development phase.

## V. CONCLUSION

This paper elaborates on the importance of considering the boundary effect in EMD-based load and renewable generation forecasting tools. Case studies show that in real-time applications, a non-tailored approach for handling the boundary effect issue can result in substantially reduced forecast performance. Further research needs to be conducted to address this issue and make the EMD-based forecast models applicable in practice.

## VI. ACKNOWLEDGEMENT

The authors gratefully acknowledge Mr. Cordell Wrishko for his insightful comments during the preparation of this study.